# Atomistic insights into Cu segregation effects on irradiation-induced defect dynamics in medium-entropy alloys

Kazi Tawseef Rahman[1], Mustofa Sakif Shahriar[1], Mashaekh Tausif Ehsan[1], Mohammad Nasim Hasan[1*]

[1]Department of Mechanical Engineering, Bangladesh University of Engineering and Technology (BUET), Dhaka-1000, Bangladesh.

[*]Corresponding author *Email address: nasim@me.buet.ac.bd*

Email: 2000tawseef@gmail.com, sakif11902117@gmail.com, mashaekh.tausif@gmail.com, nasim@me.buet.ac.bd

[*]**Corresponding Author:**

Prof. Mohammad Nasim Hasan

Department of Mechanical Engineering

Bangladesh University of Engineering and Technology (BUET)

Dhaka-1000, Bangladesh

E-mail: nasim@me.buet.ac.bd

Telephone: +8801921506445

# Atomistic insights into Cu segregation effects on irradiation-induced defect dynamics in medium-entropy alloys

**Abstract:** Copper (Cu) segregation in medium and high-entropy alloys (M/HEAs) has shown significant influence on alloy properties. In this study, we investigate the effect of Cu segregation on evolution of irradiation-induced defects in FeNiCu, a model MEA, using hybrid molecular dynamics (MD) and Monte Carlo (MC) simulations. Thermodynamically driven hybrid MC/MD annealing at low temperature resulted in a partially decomposed Cu-segregated structure (CSS) and was compared with a random solid solution (RSS) and pure Ni. Results through cumulative displacement cascades reveal that Cu-rich domains in CSS act as defect traps, accelerating interstitial-vacancy recombination and suppressing defect cluster growth. The complex potential energy landscape (PEL) in CSS disrupts dislocation propagation, leading to spatially dispersed networks. Notably, CSS exhibits reduced stair-rod dislocation density compared to RSS, highlighting its superior resistance to irradiation swelling. Localized shear strain causes dislocations to preferentially nucleate in/near Cu-rich regions though their growth is hindered by chemical heterogeneity of the alloy. Notably, prolonged irradiation induces slow Cu segregation in the RSS structure, while slowly annihilating pre-existing Cu clusters in CSS simultaneously. The findings provide atomic-scale insights into the interplay between Cu segregation and irradiation-induced defect evolution in MEAs.

# 1. Introduction

Nuclear energy is key to a sustainable future, as it reduces greenhouse gas emissions while providing a reliable energy supply to meet growing global demand [1]. In future, fourth-generation nuclear reactors will need to work in extremely corrosive environments with significant radiation doses and temperatures [2]. To withstand such harsh conditions, the next generation of nuclear materials must exhibit higher corrosion and wear resistance, dimensional stability under radiation, and improved radiation resistance. One of the most promising alternatives that meet all these criteria is medium-entropy alloys (MEAs, with three alloying elements) or high-entropy alloys (HEAs, commonly containing five or more alloying elements) denoted as M/HEA [3].

M/HEAs have gained massive attention for their exceptional qualities, including high strength [4] and hardness [5], superior thermal properties [6], radiation resistance [7], high tensile ductility [8], excellent corrosion [9] and wear resistance [10]. A recent review highlighted that reduced thermal conductivity, sluggish diffusion, and effect on defect energetics are the most commonly cited reasons for better radiation response in M/HEAs [3]. However, some of these proposed reasons remain subjects of ongoing debate. Reduced thermal conductivity in M/HEAs causes slower energy dissipation, leading to a longer thermal spike, which enables increased defect recombination [11]. That is why M/HEAs for nuclear applications have emerged as a hot topic for researchers in recent years.

Compared to elemental materials like nickel (Ni), M/HEAs exhibit superior radiation resistance. Recently in a molecular dynamics (MD) study, Li et al. revealed that CoCrCuFeNi HEA more effectively suppresses defect growth than elemental Ni [12]. Another study by Li et al. on CoNiCrFeMn HEA concluded that in contrast to Ni, CoNiCrFeMn shows fewer defects and dislocation loops due to high defect recombination as a result of having a lower difference in migration energy between interstitial and vacancy [13]. Similar behaviors were observed NiCoCrFe [14] and NiCoFe [15] systems. Experimental study by Yaykaşlı et al. on a novel HEA equiatomic composition CoCrFeNiAg further confirms these advantages, demonstrating both thermal stability and excellent radiation shielding properties [16].

Originally, M/HEAs were believed to form random solid solutions (RSSs) with homogeneous atomic distributions. However, recent research has shown that M/HEA forms local chemical ordering (LCO) and atomic segregation [17], often forming multiphase systems at lower temperatures [18]. The classic CoCrFeMnNi Cantor alloy, previously considered as a single-phase solid solution has been shown to form a secondary Cr-rich precipitated phase after prolonged annealing at certain temperatures [19, 20]. Recent MD studies on FeNiCr MEA by Liu et al. and Arkoub et al. showed significant effect of Cr segregation on defect and dislocation characteristics [21, 22]. Li et al. showed CoNiCrFeMn HEA demonstrates better radiation resistance in ordered configuration and lower resistance in the Cr-rich configuration [23]. These findings underscore the importance of atomic segregation in influencing the irradiation response of M/HEAs.

Copper (Cu) is often included in M/HEAs and has become a research hotspot because it tends to induce FCC phase in alloy systems [24]. Cu being a fast diffuser in alloys has a tendency of segregation [25]. This segregation tendency causes changes in alloys properties such as hardness, corrosion resistance and wear resistance [26–30]. This study investigates how Cu segregation affects irradiation-induced defect dynamics in MEAs, using equiatomic FeNiCu as a model system. The equiatomic composition of FeNiCu has shown to form stable FCC structure [31]. We employed a hybrid MD and Monte Carlo (MC) simulation scheme to obtain the partially decomposed Cu segregated configuration of FeNiCu. To investigate the prolonged irradiation damage and defect dynamics, MD simulations with up to 500 successive displacement cascades were carried out in both random solid solution (RSS) and Cu segregated structure (CSS). For comparison, elemental Ni was also simulated. We analyzed and compared defect and dislocation evolution across RSS, CSS, and Ni systems as the irradiation cascade progressed.

## 2. Methodology

### 2.1 Simulation Cell and Interatomic Potential

All the simulations in our current work were carried out using MD software LAMMPS, which is a widely recognized open-source MD simulator [32]. The simulation cell size was set to $30a_o \times 30a_o \times 30a_o$, where $a_o$ is the face-centered cubic (FCC) lattice constant, resulting in a system containing 108,000 atoms. The lattice constants for Ni and FeNiCu were 0.352 nm and 0.355 nm, respectively. Calculation details of equilibrium lattice constant for FeNiCu MEA can be found in **Supplementary Note 1**. The pure Ni and random configuration of FeNiCu were prepared using Atomsk software [33]. Equiatomic Fe, Ni, and Cu atoms were randomly placed on the FCC lattice sites in the RSS configuration. The CSS configuration of FeNiCu was created using hybrid MC/MD simulation to achieve an energy-minimized ordered system. The specific procedure for this is detailed in **section 2.2**. To describe the force field of the simulation system, we employed the embedded-atom method (EAM) potential developed by Tramontina et al. [34]. This interatomic potential is a modified version of Bonny et al. [35] which comes with Ziegler–Biersack–Littmark (ZBL) [36] modification for short-range interactions. This interatomic potential is specifically designed to study the Cu precipitation in FeNiCu alloy and has been successfully used to study the radiation resistance in FeCu core-shell nanoparticles under collision cascade simulation [34], making it suitable for our study. Further validation of the interatomic potential can be found in **Supplementary Note 2**. To get converged results, EAM potential by Deluigi et al. [37] was used to calculate the dumbbell migration energies of pure systems (**section 2.7**). The mathematical model for EAM potential is as follows:

$$E_i = \sum_{j \neq i} F_\beta \rho_\gamma(r_{ij}) + \frac{1}{2} \sum_{j \neq i} \varphi_{\beta\gamma}(r_{ij}) \qquad (1)$$

Here $E_i$ is the potential energy of atom $i$, $r_{ij}$ is the distance between atoms $i$ and $j$, $\varphi_{\beta\gamma}$ is a pair-wise potential function, $\rho_\gamma$ is the contribution to the electron charge density from atom $j$ of type $\gamma$ at the location of atom $i$, and $F$ is an embedding function that represents the energy required to place atom $i$ of type $\beta$ into the electron cloud. To quantify the local chemical ordering, we used the Warren–Cowley parameter [38], which is formulated as follows:

$$\alpha_{ij}^m = 1 - \frac{p_{ij}}{c_j} \qquad (2)$$

Here $\alpha_{ij}^m$ is the Warren–Cowley parameter for the m-th nearest neighbors' shell and $p_{ij}$ is the average probability of finding a $j$-type atom in the system. A negative $\alpha_{ij}^m$ suggests the tendency of $j$-type clustering in the m-th nearest neighbors' shell of an $i$-type atom, while a positive value means repulsion. In our work, we have only calculated $\alpha_{ij}^1$, which is the Warren Cowley parameter for the first nearest neighbors (1NNs) shell. From now on, we will denote $\alpha_{ij}^1$ as $\alpha_{ij}$ for simplicity. The OVITO-python modifier developed by Sheriff et al. [39] was used in our study to calculate $\alpha_{ij}$.

## 2.2 Hybrid MC/MD Simulation

The CSS configuration is obtained by annealing the RSS system using hybrid MC/MD simulation under canonical ensemble (as implemented in ref. [40] for FeNi system and in ref. [22] for FeNiCr system). On each MC step, the two different types of atoms are swapped places to slowly minimize the total energy of the system. Our RSS system was annealed at 300 K temperature (as low annealing temperature promotes segregation [41]) by performing one MC step followed by one MD step of isothermal-isobaric ensemble (NPT) with a timestep of 1 fs. This process was continued for 430 ps resulting in 215,000 MC steps and 63389 accepted swaps. On each MC step, 10 trial attempts were made and periodic boundary conditions were applied in all three dimensions of the simulation cell. **Fig. 1(a)** shows how, the system's potential energy is decreased as the annealing process proceeds. After 430 ps, we get a partially decomposed Cu segregated structure (CSS) compared to RSS configuration, which is sufficiently representative for the purposes of our study. During the process, the Cu atoms slowly segregate which is also seen in other literature [25]. Cu-rich domains are clearly seen in **Fig. 1(c)** in the CSS system. To quantify the ordering, we calculated $\alpha_{ij}$ of all the pairs and the results coincide with **Fig. 1(b)**, indicating significant ordering in the Fe-Ni and Cu-Cu pairs. The significantly negative value of Cu-Cu pair indicates Cu segregation. Whereas $\alpha_{ij}$ for RSS are all nearly zero, meaning a random mixture with no ordering. We also performed hybrid MC/MD simulations at different annealing temperatures (600K, 900K and 1200K) and found that with increasing temperature the segregation of Cu decreases (see **Supplementary Note 3**).

## 2.3 MD Displacement Cascade Simulation

To study the prolonged irradiation damage in the Ni, RSS, and CSS systems, we performed 500 consecutive displacement cascades which corresponds to a radiation dose of ~0.23 dpa (calculated based on the NRT model with a threshold displacement energy 40 eV [42]). The number of cascades has been limited to 500 because of our limited computational resources. The short simulation time between collision cascades, which is a restriction of MD, results in a dosage that is far larger than experimental settings. Before the displacement cascades were performed, each system was relaxed for 100 ps under isothermal-isobaric ensemble (NPT) at 300 K and zero pressure conditions. An atom is randomly chosen as the primary knock-on atom (PKA) and the entire cell was shifted under periodic conditions to move the PKA at the center. This was done to ensure that the PKA doesn't reach the cell boundaries. Then the PKA was given a velocity equivalent to 5 keV of kinetic energy in a random direction. Adaptive timestep was adopted during the entire cascade process. The Nose-Hoover temperature-rescaling thermostat [43, 44] was applied on the boundaries (thickness of 3.6 Å) to act as a heat sink and cool down the system to 300 K with a damping parameter 100 times the adaptive timestep. The interior of the simulation system was under microcanonical ensemble (NVE). Each cascade was run for 23,000 adaptive timesteps (around 44 ps) which was enough to cool down the entire system to 300 K. To eliminate any drifting of the system during cascade simulation, the center of mass of the entire system was constrained to its initial position. After each cascade, the simulation cell was shifted back to its original position under periodic conditions in order to allow analysis, which uses the original location of the simulation cell as reference. This process is repeated 500 times to achieve the prolonged irradiation damage. The adaptive timestep was changed between 0.0001 to 2 fs such that atomic movement was limited to 0.1 Å and atomic energy change was limited to 2.5% of PKA energy. The electron stopping effect was taken into account for the atoms with kinetic energies more than 10 eV. The electron stopping powers were calculated using the SRIM software package [45]. Other details of the simulation procedures can be found elsewhere [46, 47]. To get more statistically reliable findings, three independent simulations were run for each system using different seed values for prolonged displacement cascades. Meanwhile, to study the defect evolution with time in a single cascade event, 30 independent simulations were run for each system. Validation of our cascade simulation model can be found in **Supplementary Note 4**.

## 2.4 Defects Analysis and Visualization

To analyze the damage at the end of each cascade, the irradiated system is compared with the initial relaxed structure. All the visualization and analyses were done using the OVITO software package [48]. Wigner-Seitz cell method with Voronoi polyhedral was used to detect the point defects in the system. Polyhedra with more than one atom were identified as interstitials, and no atoms as vacancies. Defect clusters were analyzed with Cluster Analysis method with a cut-off radius equal to the second nearest neighbor distance. The dislocations were identified and analyzed with Dislocation Analysis (DXA) method.

## 2.5 Chemical Potential Calculation

To calculate vacancy and interstitial formation energy, chemical potentials of the elements in the system are required. In our work, chemical potentials were calculated using Widom-type substitution techniques [49]. We calculated the different chemical potentials of elements present in a system by randomly substituting one type of atom with another type:

$$\mu_A - \mu_B = \frac{1}{2}(E^{A \to B} - E^{B \to A}) \quad (4)$$

$$\mu_B - \mu_C = \frac{1}{2}(E^{B \to C} - E^{C \to B}) \quad (5)$$

$$\mu_C - \mu_A = \frac{1}{2}(E^{C \to A} - E^{A \to C}) \quad (6)$$

Here $\mu_A$, $\mu_B$, and $\mu_C$ are the chemical potentials of the atom types A, B, and C respectively. $E^{A \to B}$ is the energy of the system after a B-type atom is substituted with an A-type atom (similar expressions apply for the other cases). For calculation, we need another equation, which is:

$$E_o = N_A \mu_A + N_B \mu_B + N_C \mu_C \quad (7)$$

Where, $E_o$ is the initial system potential energy without any substitution and $N_A$, $N_B$, and $N_C$ are the number of atoms of type A, B, and C respectively. With these equations, we have calculated the values of $\mu_{Fe}$, $\mu_{Ni}$, and $\mu_{Cu}$ for Ni, RSS, and CSS. We used a FeNiCu system containing 864 atoms with equiatomic composition and performed 288 substitutions of each element type for all the alloys, which covers the substitution of all atoms of a certain type in the system. Further details can be found elsewhere [50, 51]. To generate an equivalent CSS structure in the 864-atom system, the same number of accepted Monte Carlo swaps per atom (~0.59) was performed as in the original 108,000-atom CSS structure, using MC/MD simulation.

## 2.6 Vacancy and [100] Interstitial Dumbbell Formation Energy Calculation

Vacancies are created by randomly removing one atom from the system and [100] Interstitial dumbbells are created by randomly replacing one atom with a pair of atoms equidistant from the lattice site in [100] direction. The formation energies are calculated by subtracting the potential energy of the perfect system from the potential energy of the defected system. The imbalance of the total atoms in the initial and final systems is compensated using the chemical potential:

$$E_f = E_d - E_o \pm \mu \quad (8)$$

Here $E_f$ is the corresponding formation energy of the defect, $E_o$ is the energy of the perfect system, $E_d$ is the energy of the defected system and $\mu$ is the chemical potential of the extra atom added or deleted. For vacancy, the chemical potential is added and for dumbbell, is it subtracted. 1000 calculations were performed for each type of point defects. Conjugate Gradient energy minimization method was used while calculating the energies.

## 2.7 Defect Migration Energy Barrier Calculations

Migration energy barriers were determined using the Climbing Image Nudged Elastic Band (CI-NEB) method [52]. Point defects were created using the method described in the previous section. The vacancies were exchanged with all the 12 1NNs, thus all the possible 12 pathways were considered for each vacancy during the vacancy migration energy calculation. A total of 10368 calculations were performed to get statistically accurate results. For interstitials, we rotated [100] dumbbells to [010] dumbbells on the {001} plane using the shift and rotate mechanism [53]. As rotating interstitial dumbbells doesn't always result in intended rotation in alloys because of high lattice distortion, the dumbbell migration energies were calculated using migration energies of pure structures and dumbbell formation energies [50]:

$$E_m^{A-B-C} = E_m^B - E_f^{A-B} + E_f^{B-C} \tag{9}$$

Here $E_m^{A-B-C}$ is energy barrier of rotating a [100] A-B dumbbell to [010] B-C dumbbell (B is the extra added interstitial atom). $E_m^B$ is the dumbbell migration energy of pure B, $E_f^{A-B}$ and $E_f^{B-C}$ are the formation energies of dumbbells A-B and B-C respectively. If the dumbbell type remains unchanged after rotation, the migration energy is simply taken as the migration energy of the pure system of the interstitial atom:

$$E_m^{A-B-A} = E_m^B \tag{10}$$

If the migration energy was negative, it was replaced by a small value (0.001 eV). Migration energies of all the 27 types of dumbbells (9 different combinations for each of the 3 types of interstitial atoms) that can be found in FeNiCu alloys, were calculated. To get converged CI-NEB results, EAM potential by Deluigi et al. [37] was used to calculate the dumbbell migration energies of pure systems. All the pure systems were considered FCC structures. For vacancy migration calculations, the small RSS and CSS structures mentioned in **section 2.5** were used. As for dumbbell migration calculations in pure systems, corresponding energy-minimized pure systems consisting of 2049 atoms were used. All the CI-NEB calculations were performed with 12 system replicas, including the initial and final states. The force tolerance convergence criterion and the spring constant were set to 0.005 eV/Å and 1 eV/Å respectively [21]. Examples of CI-NEB computed point defect migration minimum energy pathways (MEPs) can be found in **Supplementary Note 5**.

## 3. Results and Discussion:

In this study, the effects of CSS on irradiation-induced defect evolution in FeNiCu MEA were investigated using hybrid MC/MD simulations. We analyzed defect formation, dislocation dynamics, and atomic diffusion in both random and segregated configurations, with pure Ni as a reference. Key metrics such as number of Frenkel pairs, formation energy, migration energy, and dislocation density were used to assess defect dynamics across multiple displacement cascades.

### 3.1 Initiation and Development of Point Defects

Point defects like interstitials and vacancies start to develop as the PKA collides with other atoms. The evolution of point defects in a single displacement cascade for pure Ni and M/HEA can be described in four stages. The first phase is known as the ballistic phase where the number of Frenkel Pairs ($N_{FP}$) increases at a sharp rate. **Fig. 02(a)** shows that this phase lasts for ∼0.12 ps. During this period, little to no difference among Ni, RSS, and CSS is observed. Then starts the PKA energy depletion phase and lasts till 1 ps. During this phase, defect peak is observed. It is seen that for Ni the peak occurs earlier than the MEAs due to the fact that H/MEAs generally have lower thermal conductivity than pure Ni. Higher thermal conductivity in Ni enables quick transfer of energy throughout the lattice sites which leads to rapid energy dissipation resulting in a shorter energy depletion phase. Meanwhile, the more complex atomic environment in MEAs scatters the phonon more efficiently resulting in lower thermal conductivity which hinders the energy dissipation leading to a longer localized thermal energy peak. Therefore, a prolonged energy depletion phase is obtained. Furthermore, it is also observed that Ni has a lower defect peak than MEAs. The third stage, known as the recombination phase, is when the $N_{FP}$ starts to rapidly drop. In the third phase, fast atom mixing is made possible by the formation of a localized molten zone at the structure's core following the defect peak. This atomic mixing rate physically correlates with the atomic diffusivity and cohesive energy in the molten phase of the structure [22]. We have examined this atomic mixing for the three cases in **Fig. 10(a)**. The mixing is quantified by mean squared displacements (MSD) of all atoms in the system with the initial relaxed system as reference:

$$MSD(t) = \frac{1}{N}\sum_{i=1}^{N}\langle|r_i(t) - r_i(0)|^2\rangle \tag{11}$$

Here $r_i(0)$ and $r_i(t)$ are the coordinates of $i$-th atom at the initial and current time respectively. In **Fig. 10(a)**, it is clear that in all cases, the atomic mixing is strongly linear as a function of cumulative displacement cascade number. Pure Ni shows a lower mixing rate compared to MEAs. This is because pure Ni is a highly ordered, crystalline structure, leading to quick return of atoms to their equilibrium positions after displacement resulting in lower MSD. Also, the higher thermal conductivity of Ni plays a significant role here as it enables quicker heat dissipation resulting in faster solidification of the localized molten phase. So, atoms get a shorter period to diffuse. Whereas the complex atomic environment in MEA with lower thermal conductivity enables longer solidification time resulting in higher atomic diffusion. Additionally, the mixing rate of CSS is seen as lower than the RSS configuration. This indicates that the effect of CSS is significant in atomic mixing because the strong and uniform Cu-Cu bonds in Cu-rich domains supposedly decrease the overall diffusion of atoms resulting in lower atomic diffusivity in the molten phase. With time, $N_{FP}$ decreases as interstitial and vacancy defects recombine. This recombination phase lasts till ~1.6 ps. After this, a plateau stage is reached where only the surviving Frenkel Pairs remain. It is seen that pure Ni has more surviving defects than RSS and CSS, yet both have comparable amounts of residual defects. The recombination rates of the three structures are shown in **Table 1** which can be evaluated as:

$$\% \text{ Recombination} = \frac{N_{peak} - N_{end}}{N_{peak}} \times 100\% \tag{12}$$

Here $N_{peak}$ is the peak $N_{FP}$ and $N_{end}$ is the surviving $N_{FP}$ at the end of the first cascade event. We see that Ni has the lowest defect recombination rate whereas CSS and RSS have similar recombination rates.

Even though we see little to no difference between the RSS and CSS configurations in the first cascade event, the effect of ordering and segregation is prominent in prolonged displacement cascades. In **Fig. 2(b)**, it can be seen that the $N_{FP}$ linearly increases as the number of cascades ($N_C$) increases till ~35 cascades. After that, we see a non-linear behavior in defect progression. The non-linearity is caused by overlapping displacement cascades where the debris of a cascade affects the next cascade [54]. The growth of defects in the alloy slows down after linear region, but it still grows at a significant rate in Ni making MEA more irradiation resistant. We see significantly less defect growth in the CSS than in RSS after ~110 cascades. So, it is clear that ordering and Cu segregation have a significant impact on defect production in FeNiCu MEA. Detailed analysis to understand the better irradiation performance in MEA, especially in the ordered configuration, is conducted in the following sections.

## 3.2 Defect Energetics

To understand the defect characteristics and effect of CSS in FeNiCu MEA, point defect formation and migration energies are calculated. Formation energies indicate the point defect's formation difficulty. A higher formation energy indicates a higher formation difficulty and vise-versa. In pure Ni, vacancy formation energy ($E_f^v$) is constant at 1.56 eV. However, **Fig. 3(a-f)** illustrates a distribution of $E_f^v$ in RSS and CSS as a result of the structures' diverse local compositions. Compared to RSS, CSS exhibits greater $E_f^v$ for all types of vacancies, though these values remain lower than that of pure Ni. This increase in $E_f^v$ values due to ordering in CSS explains the lower $N_{peak}$ compared to RSS observed in **Fig. 2(a)** as fewer vacancies form in more thermodynamically stable structures. When comes to interstitial defects, it is seen that [100] dumbbell is the most energetically favorable interstitial defect in pure Ni compared to octahedral, tetrahedral, and crowdion interstitials [53]. So, we only calculated the energetics of [100] dumbbells in our present work for consistency. Compared to Ni, which has a constant interstitial dumbbell formation energy ($E_f^i$) of 4.63 eV, RSS and CSS have lower values for different types of dumbbells as seen in **Fig. 4(a)**. This variation of formation energies provides insight into the trends observed in the previous section. Since Ni has higher defect formation energies (for both vacancies and interstitials) than RSS and CSS, the $N_{peak}$ is lower in Ni compared to RSS and CSS alloys. In particular, CSS exhibits lower interstitial formation energies and higher vacancy formation energies compared to RSS. These results align with the findings of Arkoub and Jin [22] in FeNiCr MEA. Moreover, a significant difference is noticed between $E_f^v$ and $E_f^i$ where $E_f^v$ being considerably lower than $E_f^i$, indicating that interstitials are harder to form than vacancies. Furthermore, it is evident from the figures that in CSS, Cu-containing point defects are easier to form as both the vacancy and interstitial formation energies are lower compared to others. The average $E_f^i$ of Cu-X dumbbells (where X = Cu/Fe/Ni) is 3.22 eV being substantially lower than other dumbbell pairs. This is a result of the comparatively uniform potential energy landscape (PEL) provided by the Cu-rich segregated regions in the CSS structure, where the local lattice distortion is lower than in other areas. This promotes the easier formation of defects in these domains.

Meanwhile, formation energies explain the $N_{peak}$, they don't explain the recombination of Frenkel pairs. Therefore, migration energies are calculated. Migration energies are an indicator of point defects' diffusivity. Lower migration energy means higher diffusivity and vice-versa. Pure Ni has a constant vacancy migration energy ($E_m^v$) of 0.98 eV and a constant dumbbell migration energy ($E_m^i$) of 0.178 eV which agrees well with the values found in other computational and experimental studies [40, 51, 55]. Meanwhile, **Fig. 3(g-h)** and **Fig. 4(b)** show migration energy distributions of point defects in RSS and CSS. Again, the wide energy distribution was obtained because of the diverse localized composition in MEAs. $E_m^v$ of Fe, Ni, and Cu are 0.99 ± 0.18 eV, 1.07 ± 0.18 eV, and 0.77 ± 0.13 eV respectively in RSS and 1.09 ± 0.23 eV, 1.19 ± 0.29 eV and 0.73 ± 0.15 eV in CSS. While the migration barrier increases for Fe and Ni vacancies in CSS compared to RSS, Cu shows the opposite trend. This is because of the uniform Cu-rich segregated regions in CSS where the migration barrier gets close to that of pure Cu which is a constant 0.70 eV. Moreover, average $E_m^i$ of interstitial atoms Fe, Ni, and Cu are found to be 0.154 eV, 0.180 eV, and 0.106 eV respectively in RSS and 0.172 eV, 0.246 eV, and 0.150 eV in CSS. The overall increase in both vacancy and interstitial migration energies can be explained by the roughened PEL because of the developed ordering in the CSS system [56]. Higher diffusivity of Cu in alloys indicated by lower vacancy and interstitial migration energies is also observed in other studies [57, 58] which aligns with our findings. Individual calculated migration barrier values for all 27 types of dumbbell rotations can be found in **Supplementary Note 6**. Overall, the interstitial migration energies are found to be substantially less than that of vacancies suggesting higher diffusivity of interstitials compared to vacancies. Therefore, interstitial defects will have a higher tendency to form clusters than vacancies which will be discussed in detail in **section 3.3**.

## 3.3 Transition of Point Defects to Defect Clusters

With the cut-off radius being set as second neighbor distance, every point defect within the range was considered to be a member of the same defect cluster. The term cluster size is used to indicate the number of defects within a cluster. Clusters containing 2-30, 31-100, and 100+ point defects were defined as small, medium, and large clusters respectively for interstitial clusters. Whereas clusters containing 2-10, 11-30, and 30+ point defects were defined as small, medium, and large clusters respectively for vacancy clusters. In **Fig. 5**, the number of defects involved in a cluster are simply calculated by multiplying cluster number with cluster size. So, increase or decrease in number of defects in a cluster implies corresponding change in cluster numbers. From **Fig. 5**, it is evident that vacancy cluster growth is much more difficult than that of interstitial because of the significant difference in migration energy as discussed in the previous section. Lower migration energies of interstitials allow easier diffusion in the system and accumulate with other interstitials. Whereas the vacancies are more dispersed in the system as seen in **Fig. 7**. In **Fig. 5**, both interstitial and vacancy single clusters shows an approximate increasing trend with increasing number of cumulative cascades in all the systems where Ni shows the lowest numbers for interstitials and the opposite for vacancies. Small and medium interstitial clusters are significantly low for Ni compared to the MEAs. As for alloys, medium interstitial clusters grow steadily at first before showing a sharp decline. Large interstitial clusters are significantly high in Ni compared to alloys, which explains the significantly low numbers of small and medium clusters, as they likely accumulate together and form large clusters. Meanwhile, the large clusters in alloy systems grow at a much lower rate. The rapid fall of medium interstitial clusters in alloy systems explains the delayed growth of large clusters as higher cascade number promotes the growth of large interstitial clusters. The medium clusters combine with each other and forms the large interstitial clusters. Moreover, the number of interstitials in interstitial clusters in CSS is less than RSS in most cases of cluster size, which suggests local ordering inhibits interstitial cluster formation and growth.

As for vacancy clusters, both small and medium-sized clusters grow gradually with the increasing number of cascades in the Ni system. For alloys, small vacancy clusters also gradually increase but medium clusters show a gradual decrease in number after 300 cascades. Large vacancy clusters also grow rapidly in Ni just like interstitial clusters from the very beginning but in a significantly lower number. In contrast, the alloy systems show a much-delayed growth especially the CSS configuration where no large clusters can be seen within 400 cumulative cascades. This strongly suggests higher irradiation resistance in CSS thanks to its roughened PEL and higher migration energies. Lower large vacancy clusters in the alloy systems are a direct indicator of higher resistance to irradiation swelling which is discussed in more detail in **section 3.4**. **Fig. 6** shows the growth of the largest cluster size with increasing displacement cascades. Ni shows around twice the number of interstitial defects and around twice the number of vacancy defects in the largest cluster compared to the alloys. This demonstrates the superiority of MEAs over pure Ni in terms of irradiation resistance. While both RSS and CSS show similar growth, CSS still suppresses the aggregation of interstitial and vacancy point defects a little more than the RSS configuration. The clear details of defect growth and interaction can be seen in the **Supplementary Videos**.

To investigate the number of different constituents in the defect clusters, the percent constituents in interstitial clusters excluding single clusters as a function of cumulative displacement cascade number are plotted in **Fig. 10(c-d)** for RSS and CSS systems. In RSS, the Cu percentage doesn't change much and fluctuates around 0.28%. Whereas a gradual decrease in Ni percentage and a gradual increase in Fe percentage are seen as the irradiation dose increases. This indicates Ni interstitials are being replaced by Fe interstitials with increasing irradiation dose. This can be explained by the lower interstitial migration energies of Fe interstitials than Ni interstitials seen in **Fig. 4(b)**. Meanwhile, in CSS, the Cu percentage is significantly higher at the beginning than its overall 0.33% composition ratio in the system. This suggests that interstitial defect clusters prefer to locate in Cu-rich domains (more on this in **section 3.4**). But it rapidly decreases as more and more cascade events occur. This indicates the destruction of Cu-rich domains with increased radiation dose. As the Cu percentage decreases in the interstitial clusters, they get replaced with Ni and Fe interstitials which explains the increasing trend of Ni and Fe in **Fig. 10(d)** as the cumulative cascade number increases. Moreover, it seems that the percent constituent in CSS configuration proceeds to reach a common stable value with increasing cumulative cascades.

## 3.4 Dislocation Characteristics

Dislocation loops are formed due to the aggregating nature of point defects. When the point defects form clusters, dislocation loops start to develop around them. The evolution of dislocations can be characterized in three stages. First is nucleation then growth and lastly forming dislocation networks [21]. In our work, the dislocation loops are identified and analyzed using DXA method. The dislocation distribution in Ni, RSS, and CSS can be seen in **Fig. 7** after 100, 300, and 500 cumulative displacement cascades. From the snapshots of the systems, it is evident that dislocation forms around point defect clusters. Especially the Shockley and Frank dislocations are generally formed around interstitial clusters, while the Stair-Rod dislocations are generally formed around vacancy clusters. It can be seen that, as the number of cascades increases the dislocations also increase. For $N_c = 100$, the Shockley dislocations in RSS and CSS are very dispersed forming small-sized chained loops. While in Ni, a large-sized Shockley dislocation chain is seen. This trend is consistent in $N_c$ = 300 and 500 too. This indicates growth of dislocations in MEA is hindered by complex PEL and local chemical compositions and concluding the superiority of MEA over pure structures. Additionally, the snapshots for $N_c$ = 500 suggest the growth of Shockley dislocations is even harder in CSS configuration as the dislocations are more dispersed while RSS forms a comparatively large Shockley dislocation chain. This suggests Cu segregation in MEA has a significant impact on dislocation evolution. The dislocation loops grow by absorbing small-sized dislocations (small defect clusters getting aggregated to larger clusters) which agrees with our findings in the previous section. Moreover, it is evident that Shockley dislocation loops form complex chains which agree with other MD studies [12, 59]. Meanwhile, Shockley and Frank dislocations grow very large with consecutive cascades, the Stair-Rod dislocations don't grow that much in comparison. This is because, as mentioned earlier, Star-Rod dislocations are formed by vacancy clusters and vacancy clusters have a lower tendency of aggregation. These findings perfectly align with our results in cluster analysis.

To investigate the evolution of dislocations in greater detail, total dislocation density ($\rho_{total}$) and Stair-Rod dislocation density ($\rho_{sr}$) are plotted as a function of cumulative cascade number in **Fig. 8**. In **Fig. 8(a)**, little to no difference is observed in total dislocation density up to 35 cascades among all the systems. If we revisit **Fig. 2(b)**, no difference was also found among the system in the mentioned cascade range because of non-overlapping cascades. It suggests that similar behavior in dislocation density is also related to non-overlapping cascades. After 35 cascades, Ni shows a decrease in dislocation density growth while the previous growth is still consistent in RSS and CSS. This steep growth of dislocation density in RSS and CSS lasts till 100 cumulative cascades where no significant difference is observed between RSS and CSS in terms of dislocation growth behavior. The delayed growth in dislocation density in CSS becomes prominent after 200 cumulative cascades. After 150 cumulative cascades, the growth rate of dislocation density in MEAs becomes minimal, with CSS in particular showing fluctuations around a stable value. In contrast, Ni continues to exhibit a consistent increase in dislocation density until the end. This suggests that over time, pure Ni will develop higher dislocation densities compared to MEAs, aligning with the superior irradiation resistance of MEAs.

A major concern in materials for nuclear applications is irradiation-induced swelling [60]. This swelling is the voids in structures introduced by heavy radiation which is formed by the accumulation of vacancies. The voids are indicated by stacking fault tetrahedral (SFT) which is essentially the ratio of Stair-Rod dislocation segments to six and is an important parameter to assess the irradiation performance of materials [59]. So, to investigate the irradiation-induced swelling, Stair-Rod dislocation density ($\rho_{sr}$) is shown in **Fig. 8(b)**. Immediately it is evident that the number of SFT in pure Ni is significantly (around three times) higher than MEAs. Which suggests poor irradiation performance of pure Ni structures. Additionally, lower $\rho_{sr}$ is seen in the case of CSS compared to RSS in higher cumulative cascades. This indicates the stacking fault energy (SFE) in ordered structures is higher which agrees with the literature [61, 62]. Overall, it can safely be concluded that CSS shows better irradiation performance in terms of dislocation growth.

Moreover, the existence of CSS is supposed to cause high lattice distortion around the ordered regions resulting in significant shear strain. This high shear strain behaves as a catalyst to form dislocation loops leading the ordered regions to act as dislocation nucleation sites [62]. In our case, the prominent ordered regions are the Cu-rich segregated domains. This suggests that dislocations are likely to form in these regions. To investigate this, we have shown the snapshots of RSS and CSS systems at different radiation doses including dislocation lines in **Fig. 9**. It is clearly seen that in CSS configuration, most of the dislocations are formed inside or near the Cu-rich domains. In contrast, no preference is seen for the dislocation formation in RSS configuration. This means in CSS, defects prefer to accumulate in/near Cu-rich segregated domains. This is because these localized uniform regions facilitate the diffusion of defects, thereby acting as effective defect traps. Because outside these regions the PEL is rough thus making it energetically unfavorable for the defects to leave these regions. This trapping effect enables high defect recombination in these regions. Additionally, from the snapshots, a clear destruction of Cu-rich domains is seen as the radiation dose increases. We have quantified the level of Cu ordering using the Warren Cowley parameter for the Cu-Cu pair ($\alpha_{Cu-Cu}$). **Fig. 10(b)** vividly illustrates that $\alpha_{Cu-Cu}$ in CSS is significantly negative which means a strong Cu-Cu clustering resulting in Cu-rich segregated domains. But as the cumulative cascade increases $\alpha_{Cu-Cu}$ value increases at a decreasing rate. This clearly indicates a decrease in Cu-Cu clustering i.e., destruction of Cu-rich segregated domains due to ballistic collisions and thermal spike atomic mixing. This is why the Cu percentage in the CSS structure decreases as a function of cumulative cascade as seen in **Fig. 10(d)**. Meanwhile in RSS, $\alpha_{Cu-Cu}$ very slowly decreases with cumulative cascade. This suggests Cu is segregating at a very low rate in the random structure with increasing radiation dose. Irradiation induced Cu segregation has also been confirmed in other experimental studies [63, 64]. Moreover, the degree of order seems to reach an asymptotic/steady state value. This state may be achieved if the simulation is run for beyond 500 cascades. This behavior has also been reported in FeCr alloy under heavy irradiation [65].

## 4. Conclusion

This study investigated how Cu segregation governs irradiation-induced defect dynamics in medium-entropy alloys taking FeNiCu as a model alloy through hybrid MC/MD simulations. The partially decomposed Cu-segregated structure (CSS) showed distinct advantages over both the random solid solution (RSS) and pure Ni. The key findings reveal that Cu-rich domains in CSS act as effective defect traps that promotes enhanced interstitial-vacancy recombination while suppressing the growth of defect clusters. The complex potential energy landscape in CSS disrupts dislocation propagation, leading to more dispersed defect networks compared to RSS. Moreover, CSS demonstrated reduced Stair-Rod dislocation density which indicates superior resistance to irradiation-induced swelling. Dislocations preferentially nucleated in/near Cu-rich regions due to localized shear strain, though their growth was hindered by chemical heterogeneity of the alloy.

The study reveals dynamic ordering effects under irradiation. While CSS initially contains well-defined Cu-rich domains, prolonged irradiation gradually dissolves these clusters. Conversely, the RSS configuration shows slow irradiation-induced Cu segregation which indicates competing disordering and ordering processes. These processes are accompanied by the defect evolution, indicating coupled chemical and structural changes under irradiation.

Although the restricted irradiation dose of 500 cascades (0.23 dpa) due to limited computational resources limits observation of long-term trends, the simulations capture essential mechanisms through which atomic-scale segregation influences defect dynamics. The demonstrated enhancement of defect recombination in Cu-rich domains and inhibition of cluster growth through chemical heterogeneity elucidate how atomic segregation and chemical ordering influence radiation resistance in chemically complex alloys.

**Acknowledgement**

The authors gratefully acknowledge the high-performance computing facilities provided by the Institute of Information and Communication Technology (IICT) during this study. Additionally, the authors extend sincere thanks to Mahmudul Islam, Doctoral Student and R.A. @ MIT, USA for technical discussions at various stages of this research.

**Table 1**. Recombination rates of Ni, RSS and CSS structures

|     | $N_{peak}$ | $N_{end}$ | % Recombination |
|-----|------------|-----------|-----------------|
| Ni  | 154        | 16        | 89.61           |
| RSS | 368        | 13        | 96.47           |
| CSS | 287        | 14        | 95.12           |

# List of Figure Captions

**Fig. 1**   (a) Transient potential energy of FeNiCu system. (b) Warren-Cowley parameters calculated for the first nearest neighbor shell of the RSS and CSS configurations of FeNiCu alloy. (c) Representative configuration of pure Ni, RSS, and CSS systems.

**Fig. 2**   (a) Transient variation of number of Frenkel Pairs ($N_{FP}$), during the first cascade. (b) Variation of $N_{FP}$ versus number of cascades ($N_C$) with the shaded bands indicating standard deviations.

**Fig. 3**   (a-c) Vacancy formation energy ($E_f^v$) of Fe, Ni, and Cu respectively in RSS. (d-f) $E_f^v$ of Fe, Ni, and Cu respectively in CSS. (g) Vacancy migration energies ($E_m^v$) in RSS. (h) $E_m^v$ in CSS.

**Fig. 4**   (a) Interstitial dumbbell [100] formation energies ($E_f^i$) and (b) average interstitial dumbbell migration energies ($E_m^i$) of X-Cu-Y, X-Fe-Y and X-Ni-Y migrations (where X, Y = Cu/Fe/Ni) in RSS and CSS indicated as Cu, Fe, and Ni respectively. The total average migration energy is also shown for comparison. Constant energies for pure Ni are shown as dotted lines in both plots.

**Fig. 5**   Size evolution of point defect clusters as a function of $N_C$ in Ni, RSS, and CSS configurations ($N_{int}$: number of interstitials in interstitial clusters, $N_{vac}$: number of vacancies in vacancy clusters). (a–d) For interstitial clusters while (e–h) for vacancy clusters. The numbers in brackets in each figure represent the cluster size range.

**Fig. 6**   Size evolution of the largest clusters as a function of $N_C$ in the RSS and CSS configurations. (a) Interstitial clusters and (b) vacancy clusters.

**Fig. 7**   Snapshots of spatial distribution of point defects and dislocations in (a) RSS after 100, 300, and 500 cumulative cascades (b) CSS after 100, 300, and 500 cumulative cascades (c) Ni after 100, 300, and 500 cumulative cascades.

**Fig. 8**   The dislocation densities of total length of dislocation lines with the number of cumulative cascades in RSS and CSS. (a) All dislocation lines and (b) stair-rod dislocation.

**Fig. 9**   Snapshots of the system showing dislocation distribution with enlarged view of the circled regions in (a) RSS and (b) CSS systems.

**Fig. 10**   (a) MSD as a function of number of cascades ($N_C$) in Ni, RSS and CSS. (b) Warren Cowley parameter of Cu-Cu pair ($\alpha_{Cu-Cu}$) as a function of $N_C$ in both RSS and CSS. Percent constituents in interstitial clusters (size > 1) as a function of $N_C$ in (c) RSS and (d) CSS.

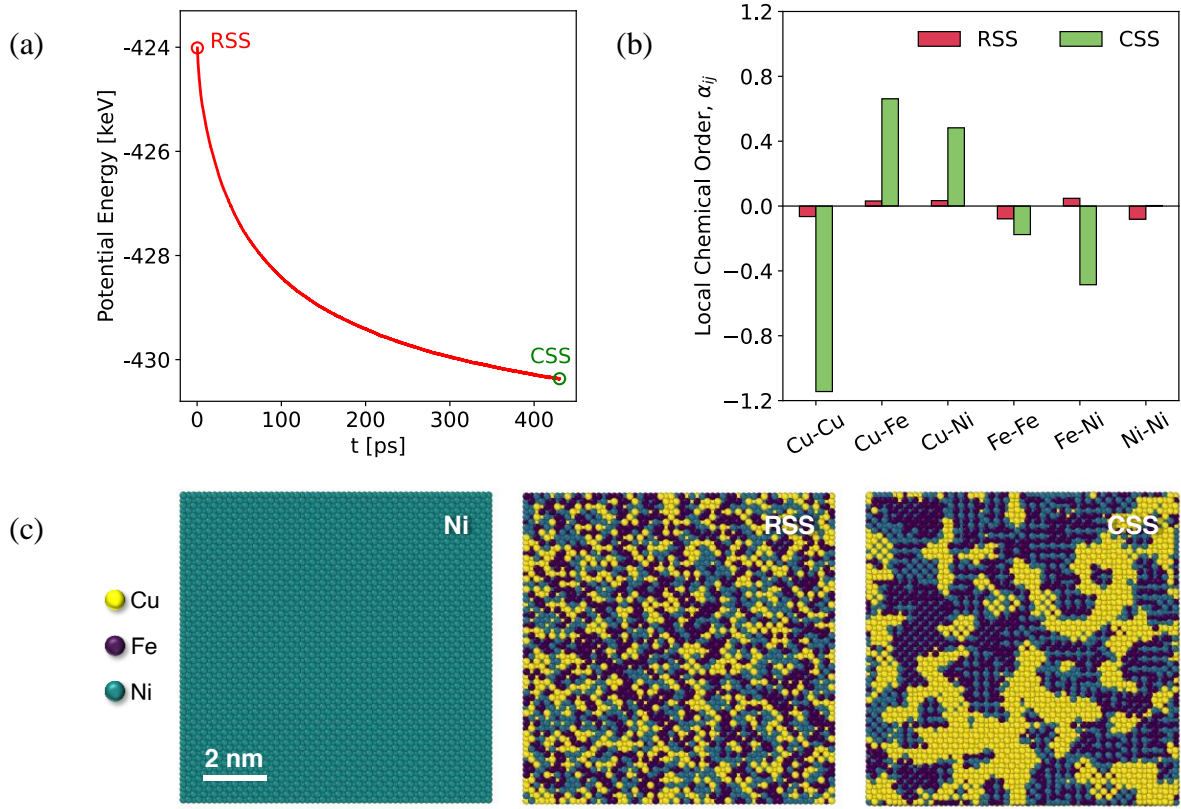

**Fig. 1.** (a) Transient potential energy of FeNiCu system. (b) Warren-Cowley parameters calculated for the first nearest neighbor shell of the RSS and CSS configurations of FeNiCu alloy. (c) Representative configuration of pure Ni, RSS, and CSS systems.

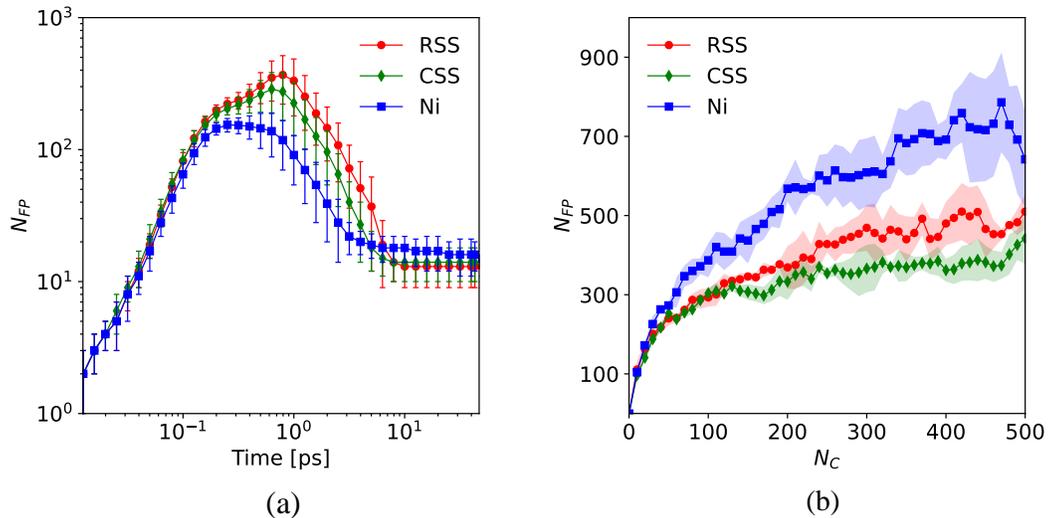

**Fig. 2.** (a) Transient variation of number of Frenkel Pairs ($N_{FP}$), during the first cascade. (b) Variation of $N_{FP}$ versus number of cascades ($N_C$) with the shaded bands indicating standard deviations.

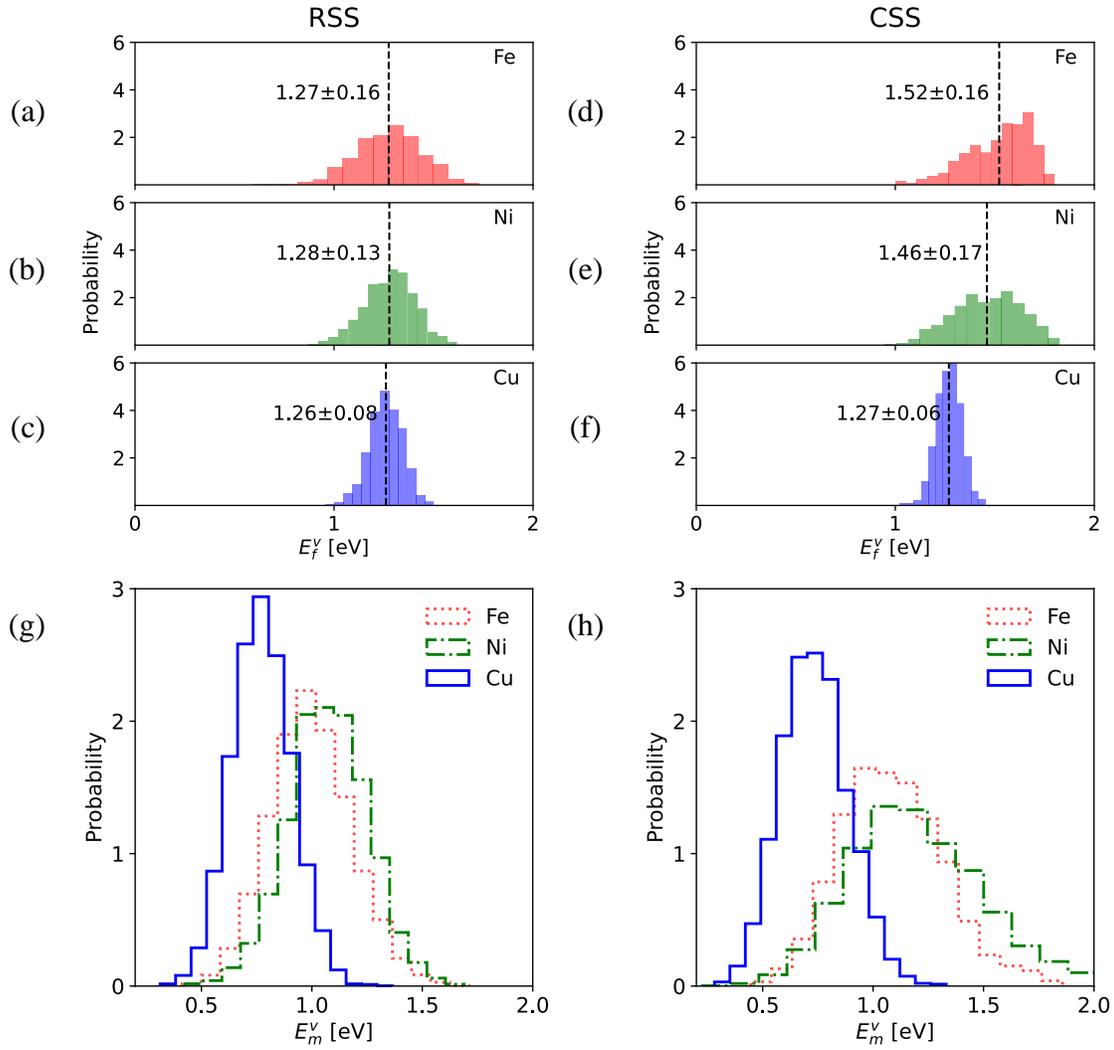

**Fig. 3.** (a-c) Vacancy formation energy ($E_f^v$) of Fe, Ni, and Cu respectively in RSS. (d-f) $E_f^v$ of Fe, Ni, and Cu respectively in CSS. (g) Vacancy migration energies ($E_m^v$) in RSS. (h) $E_m^v$ in CSS.

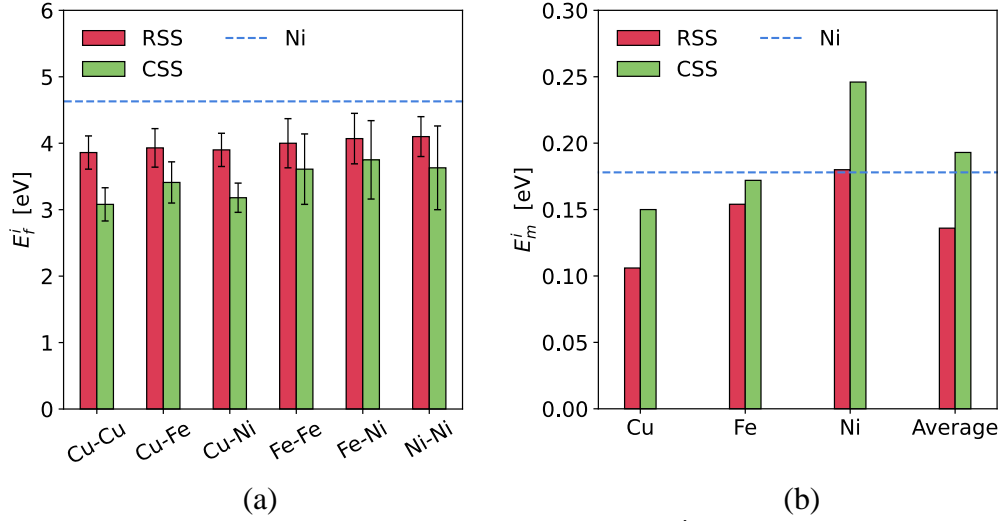

(a)                                      (b)

**Fig. 4.** (a) Interstitial dumbbell [100] formation energies ($E_f^i$) and (b) average interstitial dumbbell migration energies ($E_m^i$) of X-Cu-Y, X-Fe-Y and X-Ni-Y migrations (where X, Y = Cu/Fe/Ni) in RSS and CSS indicated as Cu, Fe, and Ni respectively. The total average migration energy is also shown for comparison. Constant energies for pure Ni are shown as dotted lines in both plots.

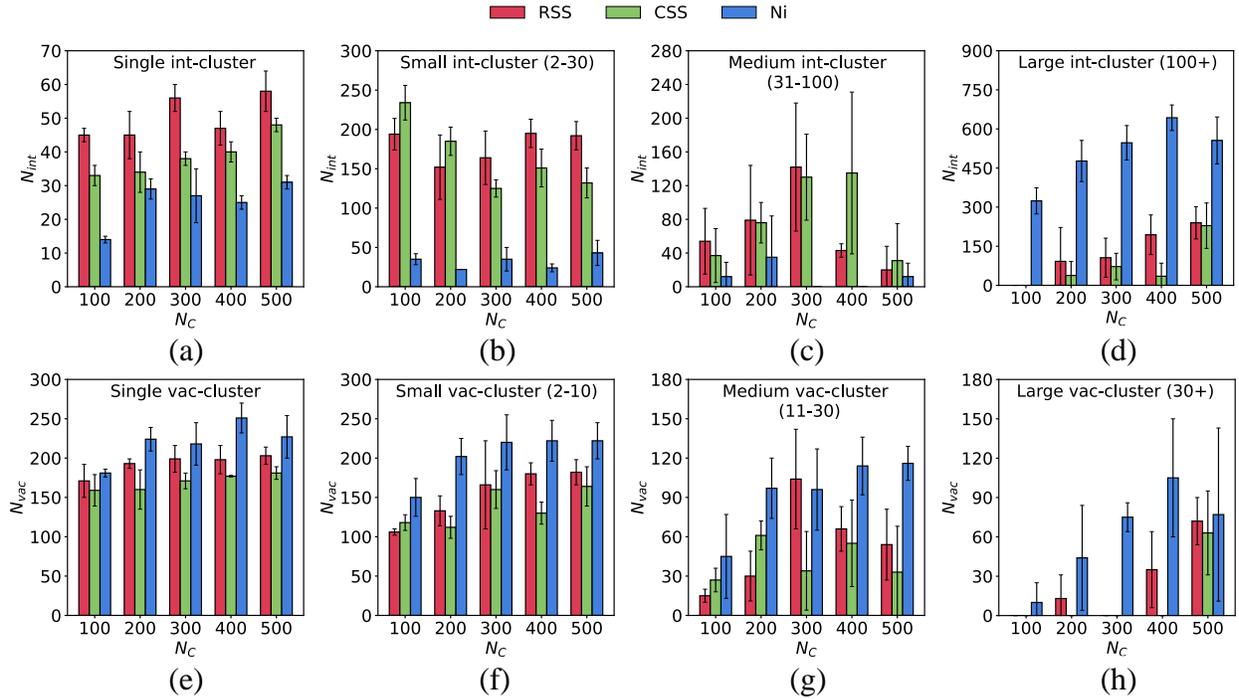

(a)             (b)             (c)             (d)

(e)             (f)             (g)             (h)

**Fig. 5.** Size evolution of point defect clusters as a function of $N_C$ in Ni, RSS, and CSS configurations ($N_{int}$: number of interstitials in interstitial clusters, $N_{vac}$: number of vacancies in vacancy clusters). (a–d) For interstitial clusters while (e–h) for vacancy clusters. The numbers in brackets in each figure represent the cluster size range.

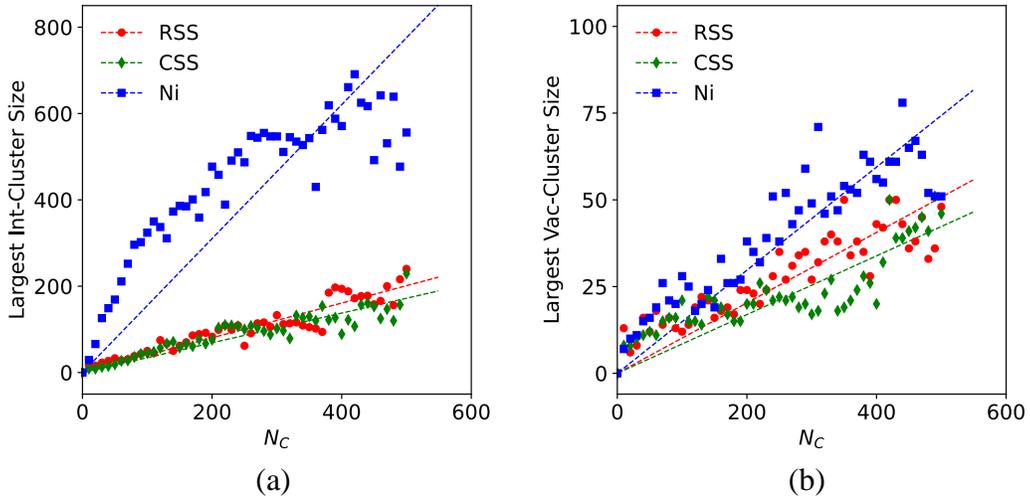

(a)

(b)

**Fig. 6.** Size evolution of the largest clusters as a function of $N_C$ in the RSS and CSS configurations. (a) Interstitial clusters and (b) vacancy clusters.

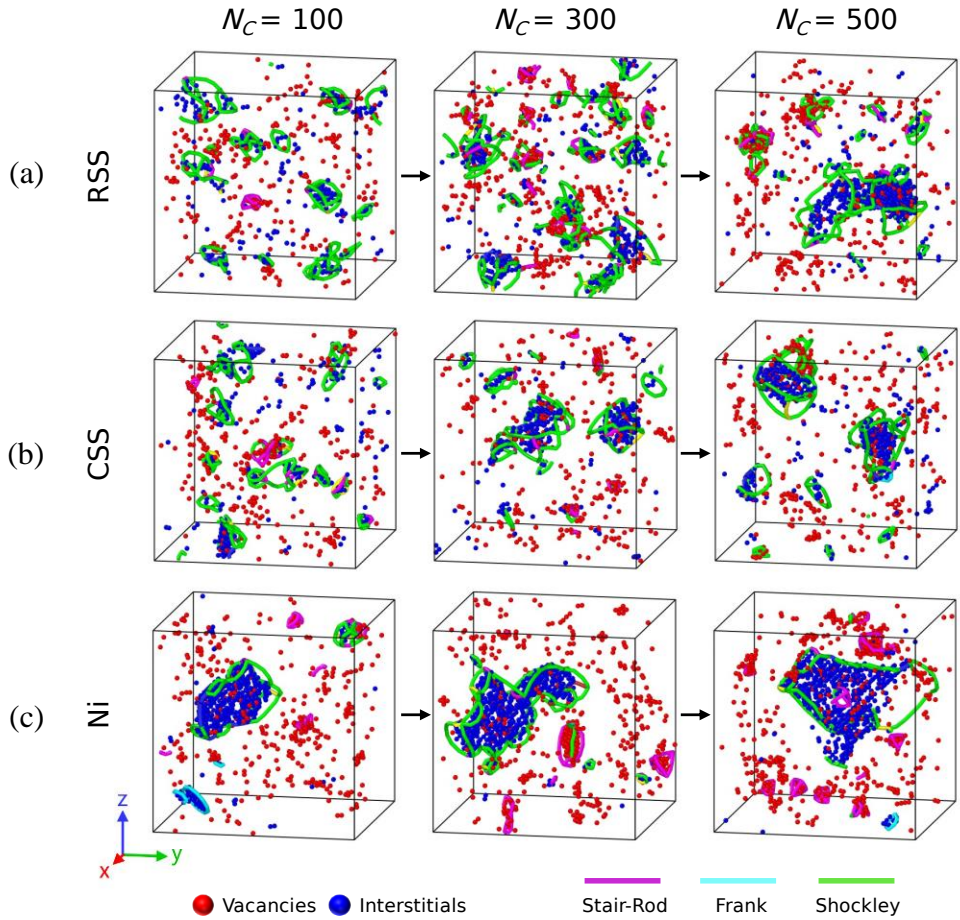

**Fig. 7.** Snapshots of spatial distribution of point defects and dislocations in (a) RSS after 100, 300, and 500 cumulative cascades (b) CSS after 100, 300, and 500 cumulative cascades (c) Ni after 100, 300, and 500 cumulative cascades.

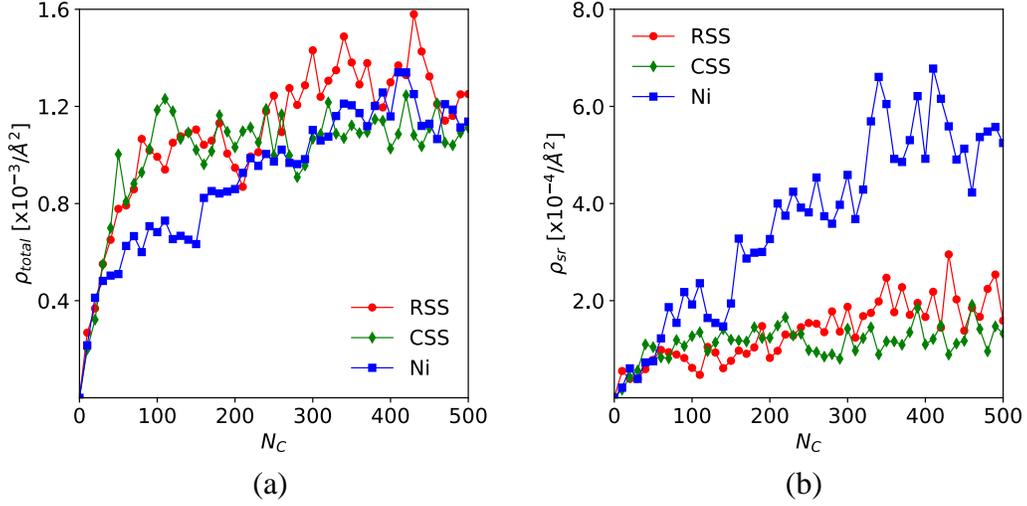

**Fig. 8.** The dislocation densities of total length of dislocation lines with the number of cumulative cascades in RSS and CSS. (a) All dislocation lines and (b) stair-rod dislocation.

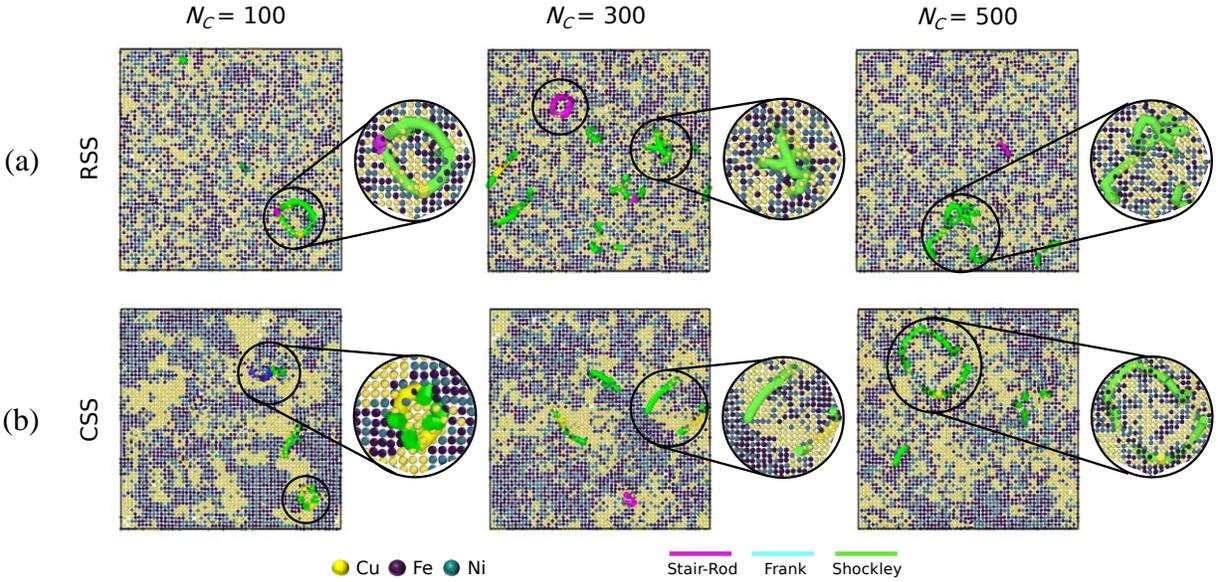

**Fig. 9.** Snapshots of the system showing dislocation distribution with enlarged view of the circled regions in (a) RSS and (b) CSS systems.

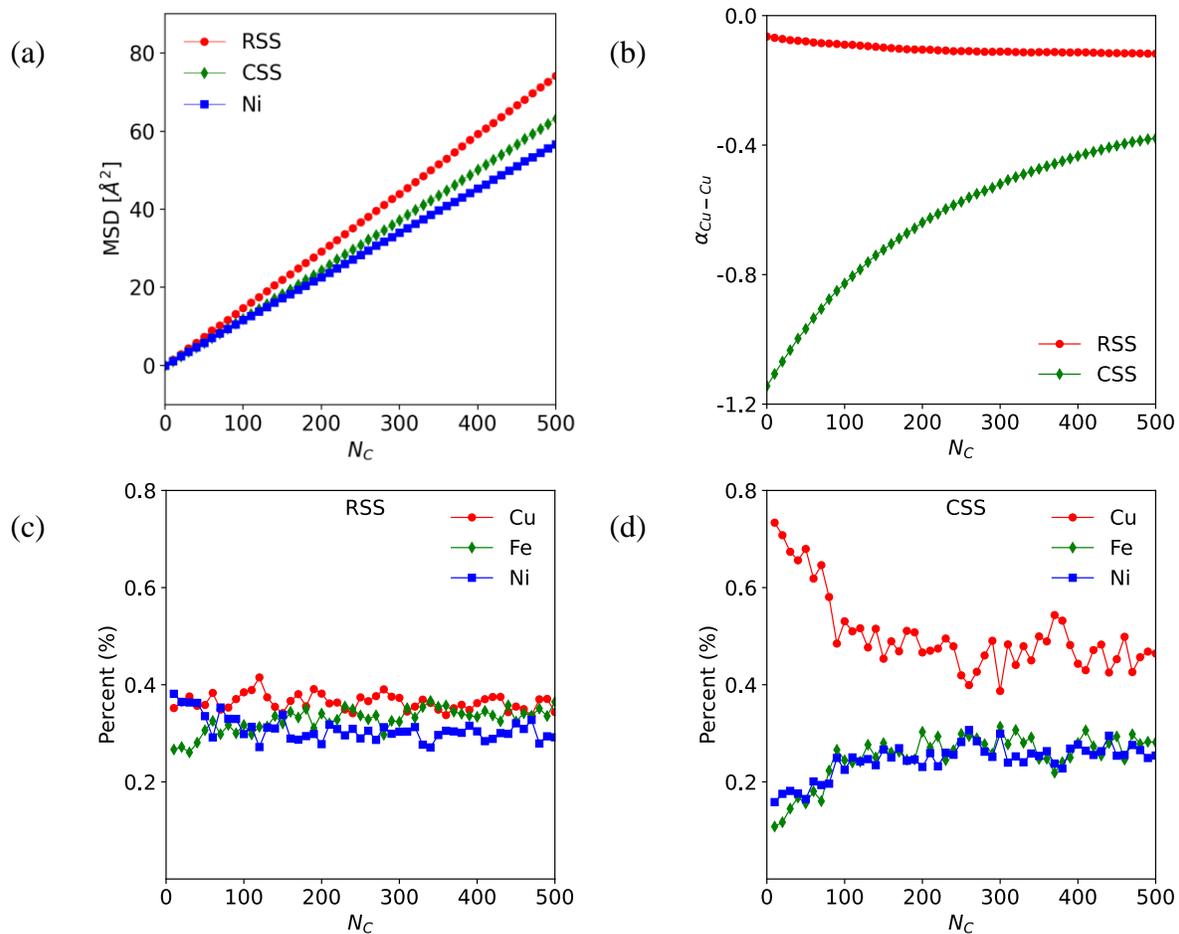

**Fig. 10.** (a) MSD as a function of number of cascades ($N_C$) in Ni, RSS and CSS. (b) Warren Cowley parameter of Cu-Cu pair ($\alpha_{Cu-Cu}$) as a function of $N_C$ in both RSS and CSS. Percent constituents in interstitial clusters (size > 1) as a function of $N_C$ in (c) RSS and (d) CSS.

Supplementary Materials to the manuscript,

# "Atomistic insights into Cu segregation effects on irradiation-induced defect dynamics in medium-entropy alloys"

Kazi Tawseef Rahman, Mustofa Sakif Shahriar, Mashaekh Tausif Ehsan, Mohammad Nasim Hasan*

## Note 1: Equilibrium Lattice Parameter in the FeNiCu MEA

Various systems of RSS were created with varying lattice constant ($a$). Then the internal energy ($E$) of each RSS system was calculated after energy minimization using the conjugate gradient method. This process was done for 10 times with different seed values to generate RSS with different atomic distribution. **Fig. S1** shows the average $E$ value as a function of $a$. The '3.55 Å' corresponding to the minimum value of $E$ was considered to be the equilibrium lattice parameter of FeNiCu MEA structure.

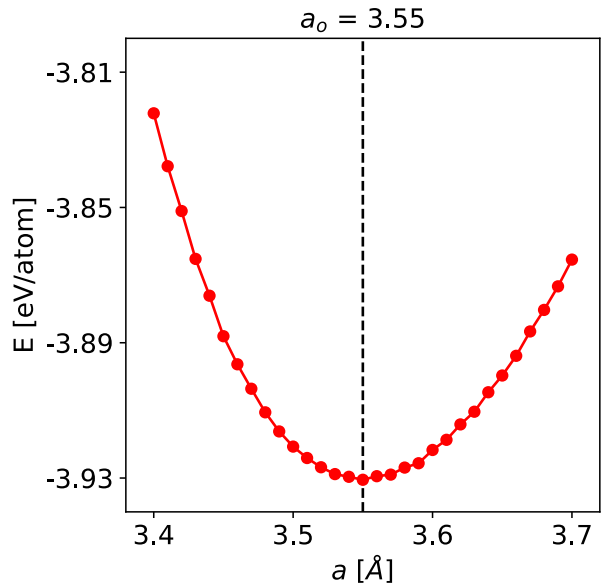

**Fig. S1.** Variation of internal energy ($E$) with respect to lattice constant ($a$), where $a_o$ is the equilibrium lattice parameter.

## Note 2: Validation of the Interatomic Potential Used

To validate the accuracy of the interatomic embedded-atom method EAM potential by Tramontina et al. [1] used in our study, we have calculated different types of defect formation energies in pure Fe, Ni and Cu systems and compared them with experimental and density functional theory (DFT) data.

**Table S1**. Defect formation energies of pure Ni

| Type of Defect | Calculated Formation Energy (eV) | Exp./DFT Data (eV) |
|---|---|---|
| Vacancy | 1.56 | 1.447 [2] |
| 100 dumbbell | 4.64 | 4.493 [2] |
| 110 dumbbell | 5.01 | 5.282 [[2] |

**Table S2**. Defect formation energies of pure Fe

| Type of Defect | Calculated Formation Energy (eV) | Exp./DFT Data (eV) |
|---|---|---|
| Vacancy | 1.756 | 1.4 ± 0.1 [4] |
| 100 dumbbell | 4.00 | 4.3-4.6 [5] |

**Table S3**. Defect formation energies of pure Cu

| Type of Defect | Calculated Formation Energy (eV) | Exp./DFT Data (eV) |
|---|---|---|
| Vacancy | 1.272 | 1.17 ± 0.07 [6] |
| 100 dumbbell | 3.15 | 3.088 [2] |
| 110 dumbbell | 3.30 | 3.386 [2] |
| 111 dumbbell | 3.59 | 3.593 [2] |

## Note 3: MC/MD Simulation at Different Annealing Temperatures

To assess the effect of annealing temperature on Cu segregation, hybrid MC/MD simulations were run at four different temperatures with same number of accepted swaps. Results show that that with increasing temperature the segregation of Cu decreases.

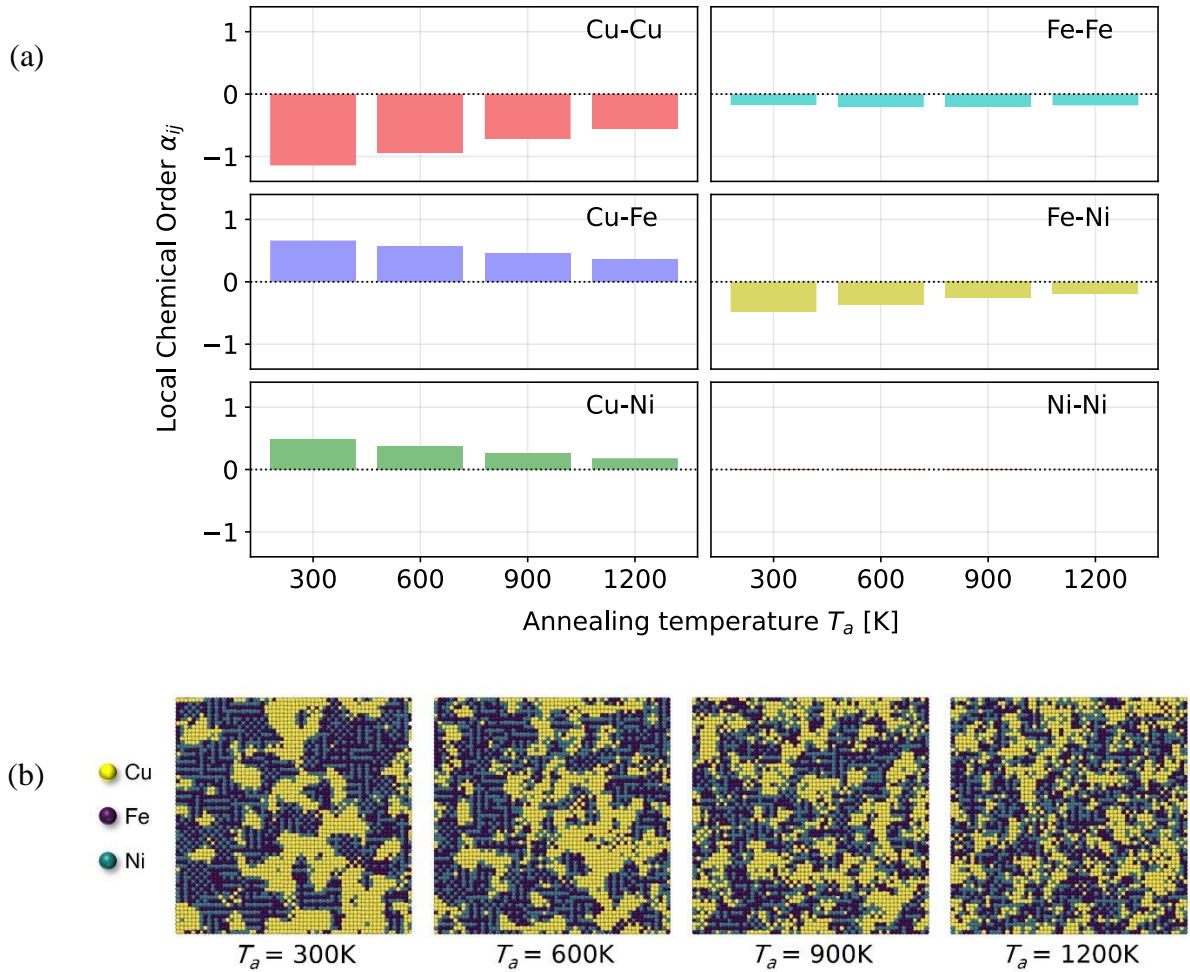

**Fig. S3.** (a) Variation of $\alpha_{ij}$ of different atom pairs in CSS with annealing temperature. (b) System snapshots of the CSS configuration at different annealing temperatures.

## Note 4: Validation of Displacement Cascade Simulation Model

To validate our displacement cascade model we ran primary displacement cascade simulation in the random FeNiCr structure with similar settings described in Arkoub et al.'s study [7]. We ran 500 consecutive cascades to evaluate our model's accuracy under prolonged irradiation. To get statistically reliable results we ran the simulation three times at random positions in the simulation cell and averaged the defect counts. From **Fig. S4**, we can see that our results closely match with the literature.

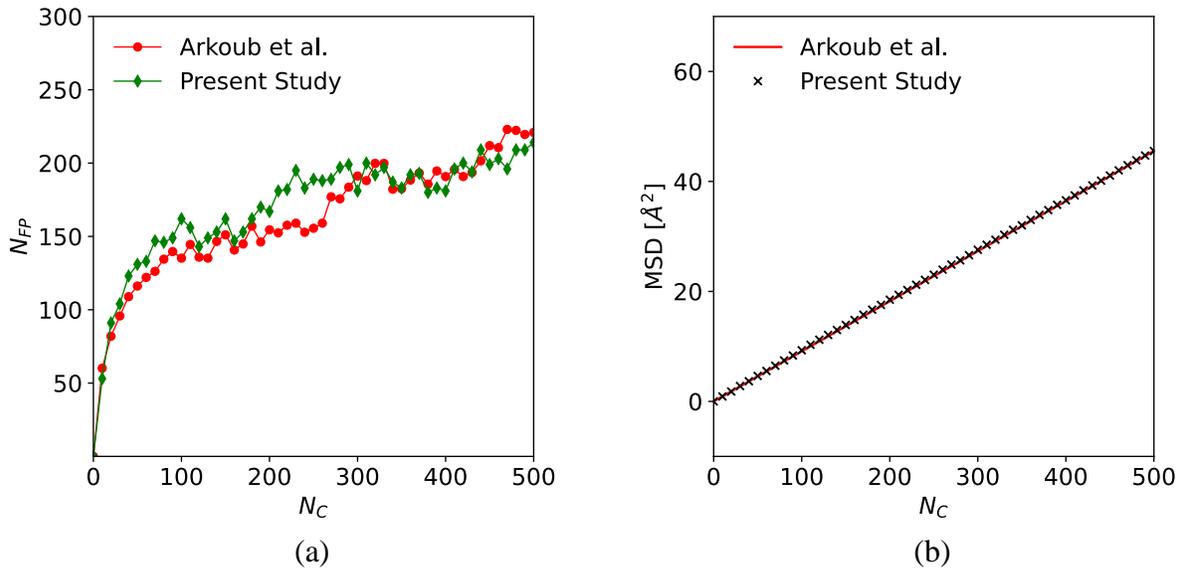

**Fig. S4.** Variation of (a) total number of Frankel Pairs ($N_{FP}$) with cascade number ($N_C$) and (b) MSD with respect to number of cascades ($N_C$).

# Note 5: Minimum Energy Pathway Calculation using CI-NEB Simulation

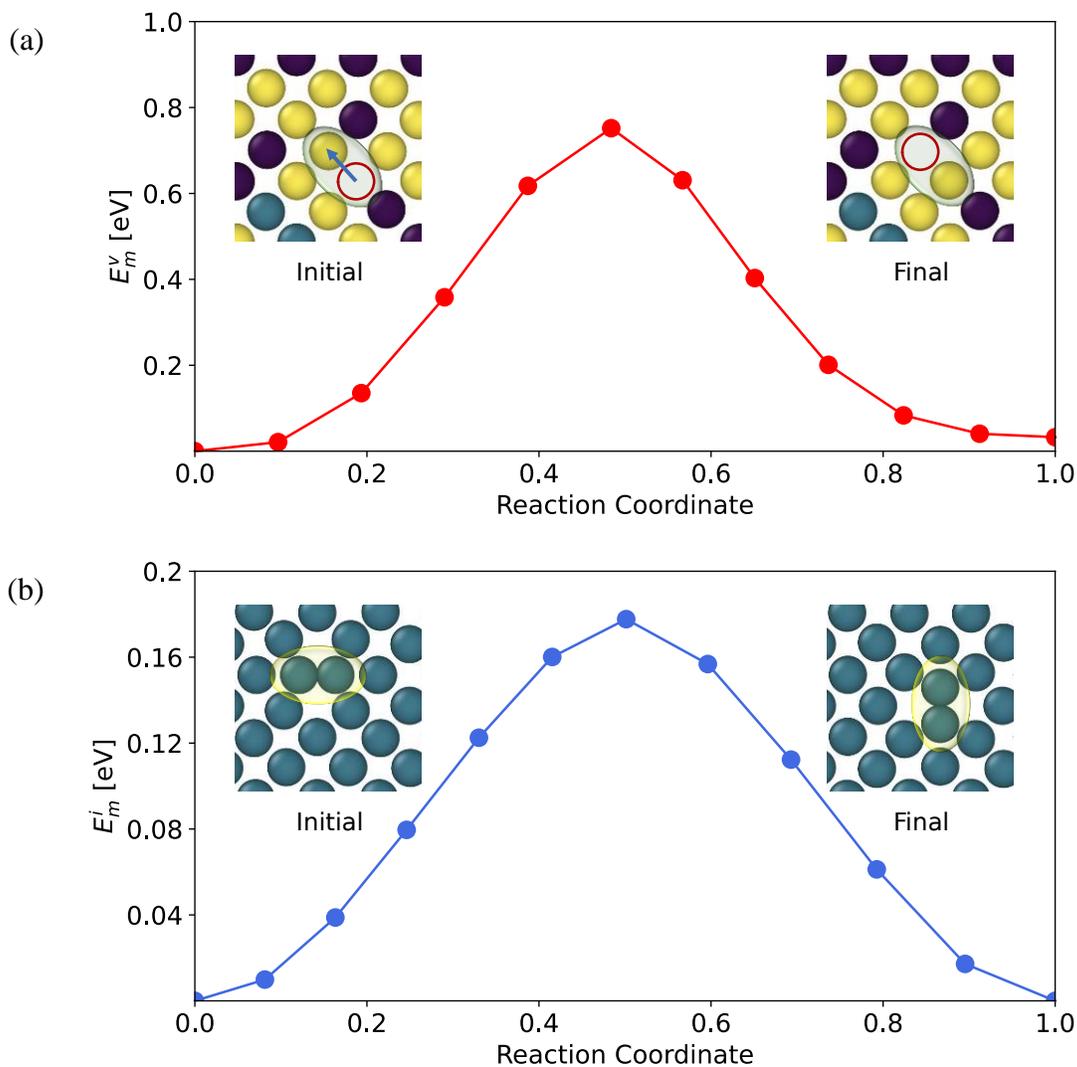

**Fig. S5.** Two examples of CI-NEB computed point defect migration minimum energy pathways (MEPs). The insets illustrate the corresponding initial and final atomic configurations. (a) The calculated MEP for a vacancy exchanging position with one of the neighbors (Cu) in RSS. (b) MEP for interstitial [100] dumbbell rotation to [010] dumbbell in pure Ni. The migration barrier is energy difference between initial state and activation state (saddle point).

## Note 6: Interstitial Dumbbell Migration Energies

The rotation types are represented as A-B-C where B is the extra added atom in the system. It means rotation of a [100] A-B dumbbell to [010] B-C dumbbell. The following tables contain the migration energies of all possible dumbbell rotations in RSS and CSS.

**Table S4**. Interstitial migration energy of all 27 types of dumbbell rotations in RSS.

| Type | $E_m^i$ | | |
|---|---|---|---|
| | X = Cu | X = Fe | X = Ni |
| Cu-X-Cu | 0.106 | 0.154 | 0.178 |
| Cu-X-Fe | 0.176 | 0.224 | 0.347 |
| Cu-X-Ni | 0.146 | 0.294 | 0.377 |
| Fe-X-Cu | 0.036 | 0.084 | 0.007 |
| Fe-X-Fe | 0.106 | 0.154 | 0.177 |
| Fe-X-Ni | 0.076 | 0.224 | 0.207 |
| Ni-X-Cu | 0.066 | 0.014 | 0.001 |
| Ni-X-Fe | 0.136 | 0.084 | 0.147 |
| Ni-X-Ni | 0.106 | 0.154 | 0.177 |

**Table S5**. Interstitial migration energy of all 27 types of dumbbell rotations in CSS.

| Type | $E_m^i$ | | |
|---|---|---|---|
| | X = Cu | X = Fe | X = Ni |
| Cu-X-Cu | 0.106 | 0.154 | 0.178 |
| Cu-X-Fe | 0.456 | 0.354 | 0.707 |
| Cu-X-Ni | 0.196 | 0.424 | 0.617 |
| Fe-X-Cu | 0.001 | 0.001 | 0.001 |
| Fe-X-Fe | 0.106 | 0.154 | 0.177 |
| Fe-X-Ni | 0.001 | 0.224 | 0.087 |
| Ni-X-Cu | 0.016 | 0.001 | 0.001 |
| Ni-X-Fe | 0.366 | 0.084 | 0.267 |
| Ni-X-Ni | 0.106 | 0.154 | 0.177 |

# Note 7: Videos of Defect Evolution

Three videos are provided to visualize the complete radiation damage process for the RSS, CSS and Ni systems respectively. Only defects and dislocations are shown and the perfect atoms are removed. Red and blue colors represent vacancies and interstitials respectively. While, magenta, sky-blue and green colors represent Stair-Rod, Frank and Shockley dislocations respectively.

RSS: "RSS_500cascades-defect-evolution.mp4"

CSS: "CSS_500cascades-defect-evolution.mp4"

Ni: "Ni_500cascades-defect-evolution.mp4"